\documentclass[fleqn,usenatbib]{mnras}

\usepackage[T1]{fontenc}

\DeclareRobustCommand{\VAN}[3]{#2}
\let\VANthebibliography\thebibliography
\def\thebibliography{\DeclareRobustCommand{\VAN}[3]{##3}\VANthebibliography}

\usepackage{graphicx}	
\usepackage{amsmath}	
\usepackage{amssymb}	
\usepackage{newtxtext,newtxmath}

\title[The limitations of $\Sigma_{1}$]{The physical connection between central stellar surface density and stellar spin in SAMI and MaNGA nearby galaxies}

\author[L. Cortese et al.]{L. Cortese,$^{1,2}$\thanks{E-mail: luca.cortese@uwa.edu.au (LC)}
A. Fraser-McKelvie,$^{1,2}$
J. Woo,$^{3}$
B. Catinella,$^{1,2}$
K. Harborne,$^{1,2}$
J. van de Sande,$^{2,4}$\newauthor
J. Bland-Hawthorn,$^{2,4}$
S. Brough,$^{2,5}$
J.~J. Bryant$^{2,4,6}$
S. Croom,$^{2,4}$
S. Sweet$^{2,7}$
\\
$^{1}$International Centre for Radio Astronomy Research, The University of Western Australia, 35 Stirling Hw, 6009 Crawley, Australia\\
$^{2}$ARC Centre of Excellence for All Sky Astrophysics in 3 Dimensions (ASTRO 3D), Australia\\
$^{3}$Department of Physics, Simon Fraser University, 8888 University Drive, Burnaby, BC V5A 1S6, Canada\\
$^{4}$Sydney Institute for Astronomy (SIfA), School of Physics, The University of Sydney, NSW 2006, Australia\\
$^{5}$School of Physics, University of New South Wales, NSW 2052, Australia\\
$^{6}$Australian Astronomical Optics, AAO-USydney, School of Physics, University of Sydney, NSW 2006, Australia\\
$^{7}$School of Mathematics and Physics, University of Queensland, Brisbane, QLD 4072, Australia
}

\date{Accepted XXX. Received YYY; in original form ZZZ}

\pubyear{2015}

\begin{document}
\label{firstpage}
\pagerange{\pageref{firstpage}--\pageref{lastpage}}
\maketitle

\begin{abstract}
The stellar surface density within the inner 1 kpc ($\Sigma_{1}$) has become a popular tool for understanding the growth of galaxies and its connection with the quenching of star formation. The emerging picture suggests that building a central dense core is a necessary condition for quenching. However, it is not clear whether changes in $\Sigma_{1}$ trace changes in stellar kinematics and the growth of dispersion-dominated bulges. In this paper, we combine imaging from the Sloan Digital Sky Survey with stellar kinematics from the Sydney-AAO Multi-object Integral-field unit (SAMI) and   
Mapping Nearby Galaxies at Apache Point Observatory (MaNGA) surveys to quantify the correlation between $\Sigma_{1}$ and the proxy for stellar spin parameter within one effective radius ($\lambda_{re}$) for 1599 nearby galaxies. We show that, on the star-forming main sequence and at fixed stellar mass, changes in $\Sigma_{1}$ are mirrored by changes in $\lambda_{re}$. While forming stars, main sequence galaxies remain rotationally-dominated systems, with their $\Sigma_{1}$ increasing but their stellar spin staying either constant or slightly increasing. The picture changes below the main sequence, where $\Sigma_{1}$ and $\lambda_{re}$ are no longer correlated. Passive systems show a narrower range of $\Sigma_{1}$, but a wider range of $\lambda_{re}$ compared to star-forming galaxies. Our results indicate that, from a structural point of view, passive galaxies are a more heterogeneous population than star-forming systems, and may have followed a variety of evolutionary paths. This also suggests that, if dispersion-dominated bulges still grow significantly at $z\sim$0, this generally takes place during, or after, the quenching phase.

\end{abstract}




\section{Introduction}
One hundred years since its conception, understanding the physical origin of the Hubble morphological sequence of galaxies \citep{hubble26} remains one of the major challenges for extragalactic astronomy. While relatively simple in its original definition, the Hubble `tuning fork' simultaneously traces changes in star formation activity (e.g., prominence of spiral arms) and stellar structure (e.g., presence of stellar bulges and/or bars), making its physical interpretation not always straightforward. Thus, it is not surprising that, throughout the decades, various re-incarnations of the Hubble tuning fork have been proposed (e.g., \citealp{devauc59,cappellari11b,kormendy12}), with an effort to provide more physical basis to the initial visual criteria used to separate different morphological types (see also \citealp{romanowsky12,cortese16,hardwick22}).     
\begin{figure*}
	\includegraphics[width=18cm]{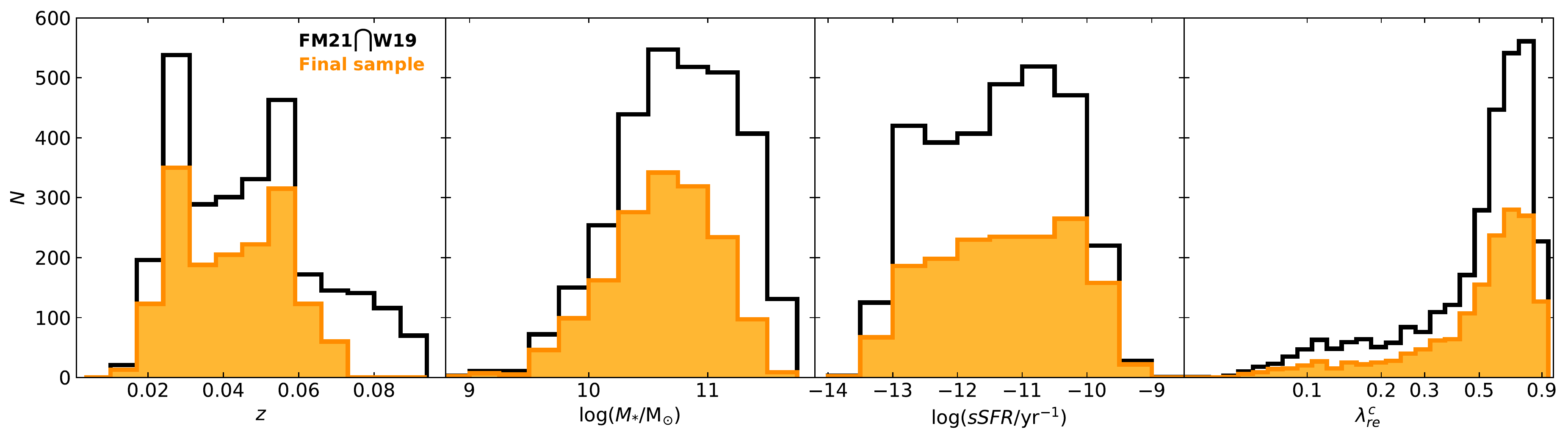}
   \caption{The distributions of (from left to right) redshift ($z$), stellar mass ($M_{*}$), specific star formation rate ($sSFR\equiv SFR/M_{*}$) and stellar spin parameter within one effective radius ($\lambda_{re}^{c}$) for our final sample (filled orange histograms). The empty black histograms show the distribution of our parent sample extracted from the intersection of the SAMI+MaNGA sample of \protect\cite{fraser21} and the $\Sigma_{1}$ SDSS catalogue of \protect\cite{woo19}.}
    \label{fig:sample}
\end{figure*}

Even when the focus is on stellar structure alone, the complexity of the galaxy ecosystem can make it difficult to link visual morphology to the actual orbital distribution of stars (e.g., \citealp{zhu18}). This is showcased, for example, by the degeneracy in the use of the term `bulge' in the context of galaxy evolution. From a physical point of view, it is generally assumed that this should be a structural component different from a disk, where most (if not all) the dynamical support comes from non circular orbits (now-a-days referred to as `classical bulge'). In reality, most works use this term to simply identify deviations in central light surface brightness/mass surface density from what is expected from an exponential disk (e.g. \citealp{kent85,peng02,simard11,lange16}). This culminates with the use of the term `pseudo-bulge' to indicate features with a disc-like 2D structure at the centre of galaxies \citep{kormendy04,erwin15}. In addition, the exact quantification of the properties of the `bulge' heavily depends on the technique used to separate this component from a disk \citep{kormendy12,meert15,cook19,tabor19,oh20}, potentially hampering our ability to unveil the evolutionary history of the inner parts of galaxies. 

An alternative approach is to refrain from trying to separate the `bulge' from the disc using 2D imaging observations and instead quantify how much, and how fast, the central core of galaxies grows. Specifically, in recent years, parts of the community have focused on the central stellar surface density of galaxies, generally measured within the central 1 kpc ($\Sigma_{1}$, \citealp{cheung12}). Being a non-parametric quantity, $\Sigma_{1}$ is independent of the actual nature of the central core of galaxies and, compared to the historically more popular stellar surface density within one effective radius, is less sensitive to changes in disk properties at larger radii, and supposed to trace the growth of the central core of galaxies \citep{fang13,barro17,walters21}.

The popularity of $\Sigma_{1}$ has further increased due to the fact that it appears to be one of the best structural parameters able to discriminate between active and passive galaxies. As initially shown by \cite{cheung12} and \cite{barro17}, and then confirmed by several others (e.g., \citealp{tacchella15,whitaker17,mosleh17,woo15,woo17,woo19,wang18,chen20,suess21}), there appears to be a clear threshold in the value of $\Sigma_{1}$ above which galaxies are almost always quiescent. This has provided indirect support to a scenario in which the growth of the central stellar core in galaxies may be physically linked to the process driving quenching (e.g., the growth of the central supermassive black hole and the onset of feedback from active galactic nuclei, AGN), a process sometimes referred to as `compaction' \citep{zolotov15,tacchella15}. While, from a theoretical point of view, compaction is expected to be more important at high redshift, recent works have suggested that it may also be an effective mechanism in the local galaxy population (e.g., \citealp{wang18}). 

While the quantification of the correlation between $\Sigma_{1}$, stellar mass, star formation rate (SFR) and other galaxy properties, as well as their dependence on redshift, has now been well established, the interpretation of the correlations in the framework of galaxy transformation, and specifically compaction, remains challenging. This is due, once again, to the fact that a higher value of $\Sigma_{1}$ at fixed other galaxy properties may be blindly assumed as evidence for the build-up of a dispersion-supported structure. This would require invoking specific physical processes able to alter the orbital structure of stars in the disc (e.g., mergers or violent disc instability, \citealp{noguchi99}). However, it has never been demonstrated whether or not $\Sigma_{1}$ is a good proxy for the presence of classical bulges in galaxies. Could it be that $\Sigma_{1}$ simply traces the stellar mass growth of discs? Is the link between $\Sigma_{1}$ and star formation quenching truly a by-product of the build-up of dispersion-supported structures in the core of galaxies?

There are very good reasons to expect that $\Sigma_{1}$ may not always trace the presence of classical bulges. Indeed, we already know that, at high stellar mass and stellar surface density, galaxies show a wide range of kinematic properties, as highlighted by the ATLAS 3D survey \citep{cappellari11}, and confirmed by more recent integral field spectroscopic (IFS) surveys \citep{cortese16,vds18,falconbarroso19,wang20,guo20}. With the advent of large IFS surveys it is paramount that we start to include kinematic information in our characterisation of galaxy structure. In particular, stellar kinematics are critical for discriminating between classical bulges and disc-like structures at the centre of galaxies. 

A wide range of stellar kinematic properties at fixed $\Sigma_{1}$ would impact on the use of the mass-SFR-$\Sigma_{1}$ plane to constrain the various paths to quiescence followed by galaxies. Indeed, the narrow range of $\Sigma_{1}$ observed in passive systems has sometimes been used to suggest a very narrow range of evolutionary paths (and thus physical processes) leading to quiescence \citep{barro17,woo19,suess21}. Of course, if this limited range was just the result of the inability of $\Sigma_{1}$ to discriminate between discs and bulges in the inner 1 kpc of galaxies, this would have major implications on our view of galaxy evolution. 

To address this issue, in this paper we present a detailed analysis of the correlation between central stellar surface density $\Sigma_{1}$ and stellar spin parameter for a sample of galaxies extracted from the overlap of the Sloan Digital Sky Survey (SDSS, \citealp{york2000}), and the Mapping Nearby Galaxies at Apache Point Observatory (MaNGA, \citealp{bundy15}) and SAMI \citep{bryant15,samidr3} IFS surveys. Our study extends on previous works that looked at the correlation between $\Sigma_{1}$ and central stellar velocity dispersion (e.g., \citealp{fang13,chen20}) by exploiting the resolved stellar kinematics maps provided by SAMI and MaNGA.

This paper is organised as follows. In Section 2, we describe our sample selection and estimates of $\Sigma_{1}$ and stellar spin, as well the ancillary data used in this work. In Section 3, we look at the correlation between $\Sigma_{1}$ and stellar spin for star-forming and passive galaxies, separately. This section includes the main results of this work. Lastly, 
we discuss the implications of our results in Section 4 and summarise in Section 5. Throughout this paper, we use a flat $\Lambda$ cold dark matter concordance cosmology: $H_{0}$= 70 km s$^{-1}$ Mpc$^{-1}$, $\Omega_{M}$=0.3 and $\Omega_{\Lambda}$=0.7.

\begin{figure}
\centering
	\includegraphics[width=7cm]{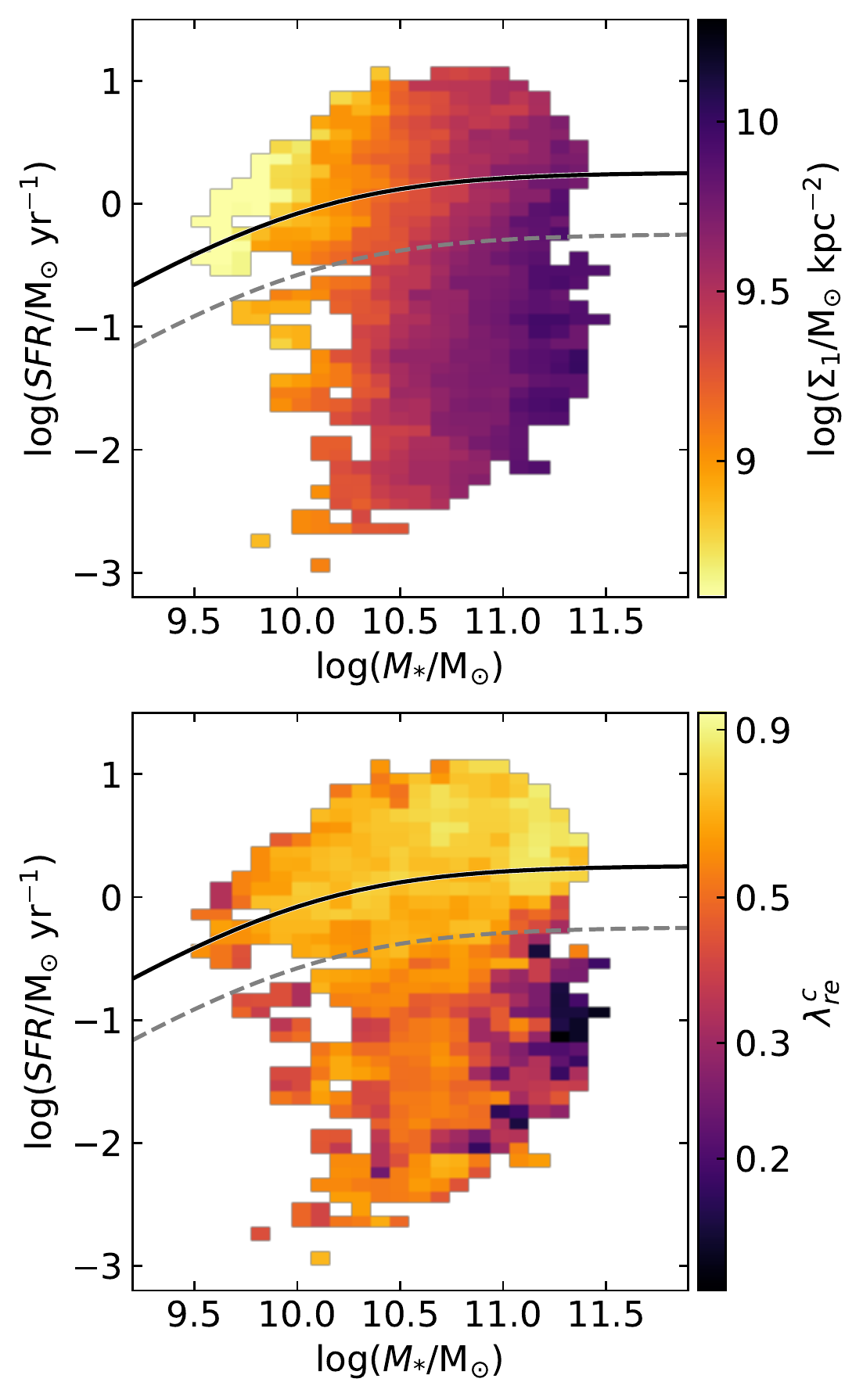}
   \caption{The SFR-M$_{*}$ plane for our sample color-coded by the median value of $\Sigma_{1}$ (top panel) and $\lambda_{re}^{c}$ (bottom panel). Medians are calculated in steps of 0.1 dex including all galaxies within 0.2 dex in both stellar mass and SFR. The solid line shows the best-fit to the star-forming main sequence of \protect\cite{fraser21}, with the dashed line indicating the threshold used here to separate star-forming and passive galaxies. Only bins with five or more galaxies are shown.} 
    \label{fig:sfr_mstar}
\end{figure}

\section{Sample selection}
Our parent sample is taken from \cite{fraser21}, who combined stellar kinematic measurements from both the SAMI and MaNGA surveys for a sample of 3289 nearby galaxies over a stellar mass range 9.5<log($M_{*}/M_{\odot}$)<12 and up to at least one effective radius. Briefly, galaxies are extracted from the overlap of the two surveys with the GALEX-SDSS-WISE Legacy Catalog (GSWLC, version 2; \citealp{salim16}) to guarantee homogeneous estimates of global stellar masses and current SFRs across the redshift range and stellar mass regimes covered by both surveys. We start from the stellar line-of-sight velocities and velocity dispersion maps produced by the SAMI \citep{vds17} and MaNGA \citep{westfall19} teams, which have been extracted from data cubes adaptively binned to a signal-to-noise of 10 using the Voronoi binning code of \cite{cappellari03}. The stellar flux-weighted spin parameter proxy within one effective radius ($\lambda_{re}$) was estimated from the line-of-sight and velocity dispersion maps following the definition presented in \cite{emsellem07}, with Petrosian $r$-band effective radii obtained from the NASA-Sloan Atlas \citep{blanton11}. In order to correct for the effect of both beam smearing and inclination we proceeded as follows. First, $\lambda_{re}$ estimates were  corrected for beam smearing following the empirical recipes presented in \cite{harborne20}, which are based on seeing and Sersic index. Beam-smearing corrected spin values were then corrected for inclination using the $r$-band ellipticity value as a proxy for inclination following \cite{delmoral20}. We point the reader to \cite{fraser21} and \cite{vandesande21} for an extensive discussion of these corrections. In the following, we use the corrected value $\lambda_{re}^{c}$ as a proxy for the stellar spin parameter, but we note that the main conclusions of this paper do not qualitatively change if observed (i.e., not corrected for either beam smearing or inclination) values are used, as discussed in Appendix~\ref{appA} of this paper. 
\begin{figure*}
	\includegraphics[width=18cm]{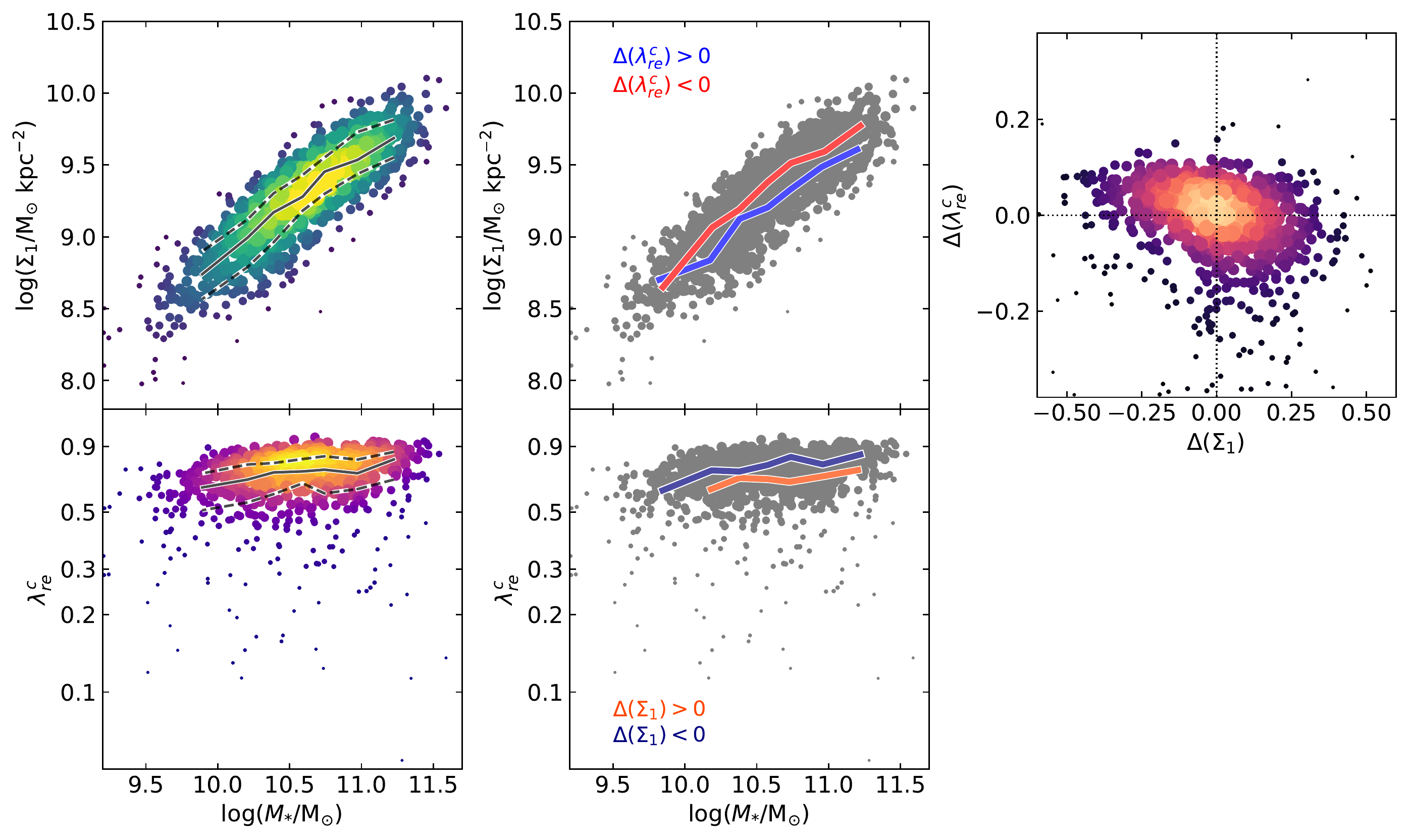}
   \caption{The correlation between $\Sigma_{1}$ and $\lambda_{re}^{c}$ on the star-forming main sequence. {\it Left column.} The $\Sigma_{1}$-$M_{*}$ (top) and $\lambda_{re}^{c}$-$M_{*}$ (bottom) for galaxies on the star-forming main sequence. Point size and color-coding reflect the probability density distributions of galaxies across each parameter space, with yellow/lighter colors indicating higher density. The black line shows the median relation for the whole sample, with dashed lines showing the 25\% and 75\% percentiles (see also Table~\ref{tab:scaling}). {\it Middle column.} Same as left column, with coloured lines now showing the running medians for galaxies with $\lambda_{re}^{c}$ (top) or $\Sigma_{1}$ (bottom) higher/lower than the observed median value at fixed stellar mass. {\it Right.} Variation in $\lambda_{re}^{c}$ as a function of variations in $\Sigma_{1}$ at fixed stellar mass. It is clear that, once we control for mass, an increase in central stellar surface density traces a decrease in stellar spin, and vice-versa.} 
    \label{fig:s1_ms}
\end{figure*}

We cross-correlated the SAMI+MaNGA sample with the measurements of central stellar surface density ($\Sigma_{1}$) presented in \cite{woo19}. These have been obtained by converting the SDSS surface brightness profiles in $g$ and $i$ band into stellar surface density profiles using the stellar mass-to-light color relation presented in \cite{fang13}. $\Sigma_{1}$ is then obtained by interpolating the stellar mass profiles and measuring the total stellar mass within 1 kpc. While $\sim$95\% of the galaxies in our IFS sample are included in the parent sample of \cite{woo19} (3102 out of 3289), for a good fraction of them 1 kpc is smaller than the point-spread function (PSF) of SDSS observations. Thus, to remove objects for which the estimate of $\Sigma_{1}$ could be unreliable and/or biased, we focus on galaxies with $z<$0.07, for which the PSF of SDSS images corresponds to less than 1 kpc, and exclude highly inclined systems (i.e., minor-to-major axis ratio $b/a<$0.5). Residual seeing effects do not affect our conclusions, as discussed in Appendix~\ref{appS1}. These cuts reduce the overlap with the IFS sample to 1599. While the sample decreases by a factor of $\sim$2 in size, we note that our final sample still spans the same parameter space in stellar mass, specific SFR and stellar spin parameter as the whole sample, as clearly shown in Fig.~\ref{fig:sample}. As expected, the redshift cut primarily affects our ability to sample the very high stellar mass regime (M$_{*}\gtrsim $10$^{11.5}$ M$_{\odot}$).

Throughout this paper, we investigate galaxies on and below the star-forming main sequence (SFMS) separately. We use the best fit to the SFMS presented by \cite{fraser21}. This was calculated using all galaxies in SAMI and MaNGA with SFRs included in the GSWLC catalog:
\begin{equation}
\label{sfms}
\log(SFR_{MS}) = 0.256-\log(1+\frac{10^{10.064}}{M_{*}})    
\end{equation}
Our active population includes all galaxies whose SFR is higher than $SFR_{MS}-$0.5 dex (see Fig.~\ref{fig:sfr_mstar}, 720 galaxies), whereas the passive population includes all galaxies with SFR below this threshold (879 galaxies).

\section{The connection between stellar central surface density and spin}
We start by showing in Fig.~\ref{fig:sfr_mstar} the SFR-M$_{*}$ plane covered by our sample, color-coded by the median value of $\Sigma_{1}$ (top panel) and $\lambda_{re}^{c}$ (bottom panel)\footnote{In this paper, we treat $\lambda_{re}^{c}$ as a log-normally distributed quantity and plot it in log scale. While this may seem at odds with what is generally done in the literature, where stellar spin is treated as a linear quantity, from a theoretical point of view the stellar spin is expected to follow a log-normal distribution (e.g., \citealp{mo98,bullok01}). 
}. We divided this plane into a grid with pixels 0.1 dex wide. For each pixel, we show the median value for all galaxies with stellar mass and SFR within 0.2 dex from the centre of each pixel. This means that we are oversampling the distribution of galaxies in this plane. 

The comparison between the two panels of Fig.~\ref{fig:sfr_mstar} already indicates that, while at fixed stellar mass passive galaxies have generally higher $\Sigma_{1}$ and lower $\lambda_{re}^{c}$ than active systems, once we focus on either star-forming or quiescent populations alone changes in $\Sigma_{1}$ may not always be followed by changes $\lambda_{re}^{c}$, and vice-versa. To better quantify the differences between the two structural indicators, in the following we examine star-forming and quiescent galaxies, separately.

\subsection{The star-forming main sequence}
In Fig.~\ref{fig:s1_ms}, we show the $\Sigma_{1}$-$M_{*}$ (top left) and $\lambda_{re}^{c}$-$M_{*}$ (bottom left) relations for all star-forming galaxies in our sample. Color-coding and symbol size scale with the probability density distributions of galaxies across both parameter spaces. The black solid line shows the running medians for both relationships, with dashed lines indicating the 25\% and 75\% interquartile ranges (see also Table~\ref{tab:scaling}).

\begin{figure*}
	\includegraphics[width=18cm]{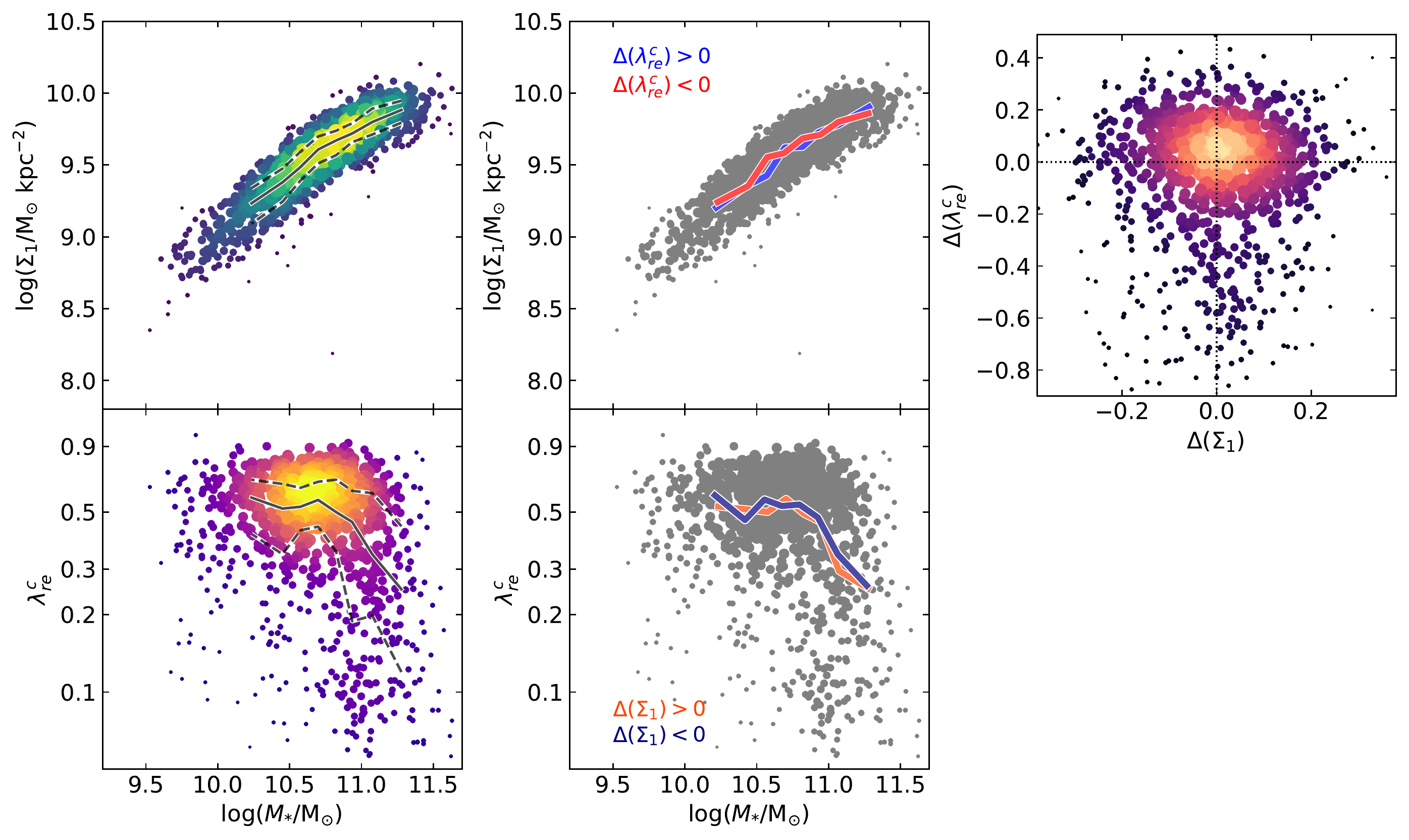}
   \caption{The correlation between $\Sigma_{1}$ and $\lambda_{re}^{c}$ below the SFMS. Colors and symbols are as in Fig.~\ref{fig:s1_ms}. Contrary to what observed in star-forming galaxies, for passive systems we find no correlation between $\Sigma_{1}$ and $\lambda_{re}^{c}$ at fixed stellar mass.}
    \label{fig:s1_noms}
\end{figure*}

As already shown by several authors in the last decade (e.g., \citealp{cheung12,fang13,suess21}), on the SFMS, $\Sigma_{1}$ monotonically increases with stellar mass. The slope of the correlation is clearly shallower than 1 implying that, if galaxies grow in mass moving {\it along} this relation (or following even shallower tracks as suggested by cosmological simulations, see e.g., \citealp{tacchella16,chen20,walters21}), most of the mass growth would take place in the outer parts of the disk (i.e., at  radii larger than 1 kpc). In other words, galaxies would not become more compact (or more centrally concentrated) while on the SFMS. This simple speculation becomes even more interesting when we look at the distribution of stellar spin parameter as a function of stellar mass. Star-forming galaxies show a narrow range of stellar spins, with $\lambda_{re}^{c}$ slightly increasing with mass (i.e., $\lesssim$0.1 dex across 1 dex in M$_{*}$; see Table~\ref{tab:scaling}), as already shown by \cite{wang20} and \cite{fraser21}. From a structural point of view, SFMS galaxies appear to form a {\it disc main sequence}.

If we combine the top and bottom left panels of Fig.~\ref{fig:s1_ms}, it appears that, on the SFMS, the fact that more massive galaxies have a higher value of $\Sigma_{1}$ {\it does not} necessarily imply that they have a more prominent dispersion-supported central component (i.e., a classical bulge). On the contrary, despite having higher central surface densities than lower mass systems, high mass galaxies are still rotationally-dominated systems from a kinematic perspective (e.g., $\lambda_{re}^{c}$>0.5). Intriguingly, this would be in line with the idea that most of the growth of mass on the SFMS at $z\sim$0 takes place beyond 1 kpc, as already hinted from the slope of the $\Sigma_{1}-M_{*}$ relation. 

Now that we have established how both $\Sigma_{1}$ and $\lambda_{re}^{c}$ vary with stellar mass, it is important to investigate whether or not these two structural parameters are correlated {\it at fixed mass}. After all, both scaling relations have a significant scatter (i.e., average 25\% to 75\% inter-quartile range IQR $\sim$0.3 dex and 0.13 dex for the $\Sigma_{1}-M_{*}$ and $\lambda_{re}^{c}$-$M_{*}$ relationships, respectively; see Table~\ref{tab:scaling}) and our goal is to establish whether higher central surface densities imply lower spins, and vice-versa. To do so, we start by quantifying the offset in $\Sigma_{1}$ and $\lambda_{re}^{c}$ for every galaxy in our sample with respect to the median value observed for SFMS galaxies at fixed stellar mass. Specifically, for each galaxy, we define $\Delta(\Sigma_{1})$ and $\Delta(\lambda_{re}^{c})$ as the logarithmic difference between the observed value and the median value of $\Sigma_{1}$ or $\lambda_{re}^{c}$ for all SFMS galaxies within 0.1 dex in stellar mass. In the top right panel of Fig.~\ref{fig:s1_ms}, we plot $\Delta(\lambda_{re}^{c})$ vs. $\Delta(\Sigma_{1})$ to show that the two quantities are clearly correlated (Spearman correlation coefficient $\rho\sim-$0.30, with a standard error from bootstrapping of 0.03). At fixed stellar mass, galaxies with higher $\Sigma_{1}$ show lower stellar spins, and vice-versa. This indicates that, once we control for stellar mass, an increase in central stellar surface density traces an increase in the importance of random motions in the gravitational support of galaxies, and thus most likely tracing the presence of small classical bulges (or potentially just the presence of a significantly thicker stellar disc). However, it is important to remind the reader that all these galaxies are rotationally-dominated, and that the decrease in spin is minimal, consistent with structural differences between those pure discs and galaxies with small bulge components (e.g., the equivalent of moving from Sd to Sb morphological types, e.g., \citealp{cortese16,vds18,falconbarroso19}).  

To show how the population of high/low $\Sigma_{1}$/$\lambda_{re}^{c}$ populates the $\Sigma_{1}-M_{*}$ and $\lambda_{re}^{c}$-$M_{*}$ scaling relations, respectively, in the middle column of Fig.~\ref{fig:s1_ms} we show the running medians for galaxies with $\Delta(\lambda_{re}^{c})>0$ and $\Delta(\lambda_{re}^{c})<0$ on the $\Sigma_{1}-M_{*}$ relation, and for galaxies with $\Delta(\Sigma_{1})>0$ and $\Delta(\Sigma_{1})<0$ on the $\Sigma_{1}-M_{*}$ relation. In both cases, galaxies with higher spin/ lower central surface densities (and vice-versa) map each other very well, and they follow parallel relations as a function of stellar mass. This confirms that the two observables are tracing the same structural changes {\it at fixed stellar mass}.

\subsection{The passive population}
The natural next step is to investigate whether or not the correlation between $\Sigma_{1}$ and stellar spin holds for galaxies that are no longer on the SFMS. A rapid glimpse at Fig.~\ref{fig:s1_noms}, which provides the same parametrisation presented in Fig.~\ref{fig:s1_ms} but for passive galaxies, immediately shows that here the situation is dramatically different. The picture put forward in the subsection above may no longer be valid below the SFMS. 

Starting from the $\Sigma_{1}$-$M_{*}$ (top left) and $\lambda_{re}^{c}$-$M_{*}$ (bottom left) scaling relations, we see significant differences. First, as already known (e.g., \citealp{fang13}), at fixed stellar mass the range of $\Sigma_{1}$ observed in passive galaxies is significantly narrower (IQR$\sim$0.19 dex) than that of star-forming systems. Second, and most importantly, the opposite happens for the stellar spin. The spread in $\lambda_{re}^{c}$ at fixed mass increases by more than a factor of 2 (IQR increasing with mass from $\sim$0.2 to 0.57 dex, see Table~\ref{tab:scaling}) and, if any trend is present, now spin clearly decreases with increasing stellar mass. However, this should not be interpreted as passive galaxies simultaneously growing their central surface density while losing their degree of rotational support. Indeed, as it is unlikely that passive galaxies grow significantly in mass via star formation, their mass increase must happen via mergers. The chaotic nature of these processes, in particular 
in the case of major mergers, makes it unlikely that the scaling relations shown in the left column of Fig.~\ref{fig:s1_noms} trace evolutionary tracks for passive galaxies. 

The lack of a correlation between central surface density and spin becomes even more striking if we compare 
$\Delta(\Sigma_{1})$ to $\Delta(\lambda_{re}^{c})$. The only difference here is that the offsets are measured from the median values of the passive population at fixed mass. As shown in the right panel of Fig.~\ref{fig:s1_noms}, $\Delta(\Sigma_{1})$ and $\Delta(\lambda_{re}^{c})$ are no longer correlated ($\rho\sim-$0.04$\pm$0.03). Thus, there is no difference in average spin between galaxies with high and low stellar central surface density, and no difference in $\Sigma_{1}$ for galaxies with low and high spin at fixed mass. In other words, in the passive population, central surface density is not a good proxy for the kinematic state of the stellar population and, specifically, to gauge the balance between rotationally- and dispersion-supported structures within one effective radius. This result is not only driven by the presence of slow rotators, but it still holds if we focus on galaxies with $\lambda_{re}^{c}>$0.25 ($\rho\sim-$0.09$\pm$0.04). Only once we focus our attention on galaxies with $\lambda_{re}^{c}>$0.4, hints of a correlation between  $\Sigma_{1}$ and $\lambda_{re}^{c}$ start to re-emerge, but still with lower statistical significance than that observed for SFMS galaxies ($\rho\sim-$0.22$\pm$0.03).

In retrospect, and as already hinted in the introduction, this result should not come as a complete surprise. Already from a visual morphology point of view we know that, while star-forming galaxies are primarily disks, passive systems include pure discs, discs plus `bulges' and pure `bulges', with `bulges' including both classical and pseudo-bulge structures. However, this morphological spread is not encapsulated by parameters that are sensitive to the 2D projected distribution of stars in galaxies, while they are clearly traced by stellar kinematic information such as the stellar spin parameter (e.g., \citealp{cappellari11,krajnovic13,cortese16}).

\section{Discussion}
The main result of this work is the lack of a direct connection between central stellar surface density and stellar spin parameter across the entire SFR-M$_{*}$ parameter space covered by nearby galaxies. For star-forming galaxies, differences in $\Sigma_{1}$, at fixed mass trace changes in stellar kinematics. This correlation disappears below the SFMS. Here, $\Sigma_{1}$ cannot be used to infer the presence (and the importance) of classical bulges in passive galaxies. We can use these findings, combined with previous works, to improve our current understanding of how galaxies evolve from a structural point of view across the SFR-M$_{*}$ plane.

Starting from Fig.~\ref{fig:s1_noms}, the striking difference in dynamic range between $\Sigma_{1}$ and $\lambda_{re}^{c}$ at fixed stellar mass and the wide range (0.2-0.57 dex) in observed stellar spin imply that the passive population spans a variety of structural properties (and hence morphologies), significantly larger than what is observed in the star-forming galaxy population. Assuming that at higher redshift SFMS galaxies still show a small scatter in stellar spin at fixed stellar mass, our finding implies that during (or after) the quenching phase galaxies can experience a wide range of structural transformation, from remaining rotationally-dominated with relatively small classical bulges at their centre, up to becoming slow-rotators fully dominated by random motions, with this extreme case becoming relevant only at high stellar masses ($>$10$^{10.5}$ M$_{\odot}$, e.g., \citealp{guo20,vandesande21}). 

It would be tempting to use our findings to question the notion of a single (or small number) of evolutionary paths towards quiescence for galaxies at $z\sim0$ (e.g., \citealp{woo19,suess21}), and instead support a more complex picture with a diversity of paths leading to quenching and structural transformation \citep{cortese09,fraser18,janowiecki20,dawesreview,saintonge22}, which would also be in line with the diversity observed in stellar population properties of passive galaxies both in the local Universe and at higher redshift (e.g., \citealp{thomas05,graves09,tacchella22}). However, we cannot exclude that just mergers alone (i.e., minor and major mergers combined)  may still explain the stellar kinematics properties of galaxies, with the wide range of spin simply due to the huge  variety of, e.g., minor mergers that a galaxy can experience (see also \citealp{grand2017,garrison18}), combined with the inevitable mass loss and associated adiabatic expansion (e.g., \citealp{vandokkum14}). Nevertheless, our results clearly demonstrate that the $\Sigma_{1}$-$M_{*}$ parameter space alone cannot be used to fully unveil the structural evolution of galaxies, and at least confirm that major/disruptive mergers become a significant pathway for galaxy transformation only in the very high stellar mass regime, where a statistically significant population of slow rotating galaxies starts to emerge.


Whether the structural transformation happens during or after quenching \citep{cortese19}, and if the same physical process is responsible for both, cannot be established from this analysis. However, a comparison between the range of $\Sigma_{1}$ and $\lambda_{re}^{c}$ covered by the active and passive population, as shown in Figs.~\ref{fig:s1_ms} and \ref{fig:s1_noms}, suggests that the bulk of dispersion-supported cores in our sample do not form on the SFMS. In other words, we are not seeing major classical bulge growth on the SFMS, before quenching. Of course, there is a small fraction of star-forming galaxies that shows spin parameters consistent with a high degree of random motion (i.e., $\sim$10\% with $\lambda_{re}^{c}$<0.4) but, as already shown by \cite{fraser21}, these are primarily interacting systems and cannot, alone, explain the observed difference between the active and passive populations. Similarly, while the scatter in stellar spin observed in our SFMS sample indicates that galaxies may be growing dispersion-supported central structures in the SFMS, these remain a relatively small fraction of the total galaxy mass, and correspond to the typical population of early-type star-forming spirals observed in the local Universe (e.g., \citealp{cook19,falconbarroso19}). This is even more the case if we take into account the fact that small changes in $\lambda_{re}^{c}$ in the SFMS may also trace the presence of a thick disc component, and not of a central dispersion-supported bulge, suggesting that our findings provide some conservative upper-limits on the importance of dispersion-supported bulges in the SFMS. 

As discussed by \cite{croom21} and \cite{cortese19}, part of the differences in stellar spin between the active and passive populations could be due to a combination of size evolution and progenitor bias, as most of the quiescent systems in our sample left the SFMS a few billion years ago, at the very least. However, even ignoring this, we do not really see a large family of galaxies with kinematic properties consistent with the presence of prominent central massive classical bulges (e.g., $\lambda_{re}^{c}<$0.5) on the SFMS (see also \citealp{morselli17,cook20}). Either most of classical bulges were formed at earlier epochs, via processes that have gradually become less efficient in recent times, or their mass growth takes place when (or after) galaxies have started leaving the SFMS. This is a key difference between using $\Sigma_{1}$ and $\lambda_{re}^{c}$ as an indicator of structural transformation: i.e., while, when observed in two dimensions, all passive systems have similar central stellar surface densities consistent with that expected for classical bulge-dominated systems, from a stellar kinematic point of view they are far from being a homogeneous family, in particular for stellar masses greater than $\sim$10$^{10.5}$ M$_{\odot}$. An increase in central stellar surface density cannot be blindly interpreted as evidence for an increase in the importance of random motions on the gravitational support of galaxies. Indeed, as shown in Fig.~\ref{fig:s1_ms}, on the SFMS, galaxies grow in both mass and central stellar surface density \citep{walters21}, but their stellar spin either remains constant or even slightly increases. 

Excitingly, the potential disconnect between quenching and morphological transformation as well as the large variety of merger-driven kinematic changes in galaxies suggested here appear in line with predictions from large hydro-dynamical cosmological simulations such as EAGLE (e.g., \citealp{lagos18,lagos22}) IllustrisTNG (e.g., \citealp{tacchella19,park21}). However, it is important to note that some of these findings may be affected by spurious collisional heating \citep{ludlow21} and next generation/higher resolution runs are needed to fully confirm these scenarios and allow us to perform a quantitative comparison between observations and simulations (see also \citealp{vds19}).

The main limitation of our work is that we are not comparing stellar surface density and stellar spin for the same regions. While $\Sigma_{1}$ traces the inner 1 kpc, $\lambda_{re}^{c}$ is measured within one effective radius, thus encapsulating a larger area (on average a factor of $\sim$3 larger in radius for passive systems and $\sim$4 for star-forming ones) and potentially more prone to trace changes in the outer parts of galaxies. Unfortunately, with current facilities, it is practically impossible to obtain higher spatial resolution IFS observations for $\sim$1500 galaxies to quantify $\lambda_{re}^{c}$ within 1 kpc. Thus, it is important to discuss if, and how, our conclusions may be affected by this issue. 

On the SFMS, it is possible that the correlation between spin and central surface density becomes even stronger, but the main conclusion of star-forming galaxies remaining rotationally-dominated systems and having only small classical bulges would hold. Below the SFMS, it is difficult to think of a scenario in which the lack of correlation between $\Sigma_{1}$ and $\lambda_{re}^{c}$ would only be due to the different apertures used to estimate the two quantities. This could only happen if the bulk of passive rotationally-dominated galaxies becomes consistent with being a slow rotator within 1 kpc. While a smaller aperture does imply a smaller rotational velocity and thus a narrower range of $\lambda$ (assuming a flat velocity dispersion profile), it is unlikely that the scatter in spin at fixed mass decreases so much to become consistent, or even smaller, than that observed in the SFMS. Indeed, from previous studies of the radial cumulative distribution of spin parameter in early type galaxies, we know that the variety in spin parameter values is already present in the very inner parts of galaxies (e.g., \citealp{emsellem11,arnold14,vds17}). 

What we have been able to test is whether our results hold if we use the stellar surface density within one effective radius ($\Sigma_{e}$ instead of $\Sigma_{1}$ as a proxy for 2D structure). While, as already known, the scatter in the $\Sigma_{e}$-M$_{*}$ relation is larger than what observed for the $\Sigma_{1}$-M$_{*}$ trend, we still find that $\Sigma_{e}$ and  $\lambda_{re}^{c}$ trace each other on the SFMS while they are uncorrelated in the passive population.   

Thus, despite the fact that the different apertures used in this work may affect the detailed quantification of the correlation between $\Sigma_{1}$ and $\lambda_{re}^{c}$, we expect that our main conclusion should qualitatively hold once measuring the spin parameter within 1 kpc for large statistical samples will become possible.



\section{Summary}
In this paper, we have combined estimates of the 2D  stellar surface density ($\Sigma_{1}$) of galaxies within their central 1 kpc, with measurements of their stellar spin parameter within one effective radius ($\lambda_{re}^{c}$), to determine if changes in $\Sigma_{1}$ map variations in stellar kinematics. We showed that:
\begin{itemize}
    \item On the SFMS, at fixed stellar mass, galaxies with higher $\Sigma_{1}$ have lower spin, and vice-versa, but the vast majority of star-forming galaxies are rotationally-dominated (i.e., discs) systems. While $\Sigma_{1}$ clearly grows with increasing stellar mass, the stellar spin either remains constant or increases by only $\lesssim$0.1 dex over 1 dex in stellar mass. 
     \item Below the SFMS, $\Sigma_{1}$ and $\lambda_{re}^{c}$ are no longer correlated. Passive galaxies show a narrow range of $\Sigma_{1}$ but a wide range of $\lambda_{re}^{c}$ values. Here, while $\Sigma_{1}$ still increases with increasing mass, $\lambda_{re}^{c}$ decreases with mass, although with a large scatter.   
\end{itemize}
Together, our findings imply that, in passive galaxies, the central stellar surface density cannot be used to disentangle rotationally- from dispersion-supported structures. 
In the context of galaxy evolution and structural transformation at $z\sim$0, we interpret our results as follows:
\begin{itemize}
    \item Only small classical bulges are able to grow while galaxies are on the SFMS. Major {\it compaction} episodes where dispersion-supported structures grow and create bulge-dominated star-forming systems are extremely rare at $z\sim$0.
    \item If classical bulge formation is still efficient at $z\sim$0, this takes place primarily during (or after) galaxies have started their quenching phase. 
    \item Passive galaxies are, from a structural point of view, an heterogeneous population. While this may still be consistent with a two-way path towards quiescence, at this stage it cannot be excluded that a multitude of evolutionary paths (and potentially physical processes) are responsible for the transformation of active discs into quiescent systems. 
\end{itemize}

\section*{Acknowledgements}
We thank the anonymous referee for useful comments which improved the quality of this manuscript.
This research was conducted by the Australian Research Council Centre of Excellence for All Sky Astrophysics in 3 Dimensions (ASTRO 3D), through project number CE170100013. LC acknowledges support from the Australian Research Council Discovery Project and Future Fellowship funding schemes (DP210100337, FT180100066). JvdS acknowledges support of an Australian Research Council Discovery Early Career Research Award (project number DE200100461) funded by the Australian Government. JBH is supported by an ARC Laureate Fellowship FL140100278. JJB acknowledges support of an Australian Research Council Future Fellowship (FT180100231). SMS acknowledges funding from the Australian Research Council (DE220100003). Parts of this work were performed on the OzSTAR national facility at Swinburne University of Technology. The OzSTAR program receives funding in part from the Astronomy National Collaborative Research Infrastructure Strategy (NCRIS) allocation provided by the Australian Government. 

The SAMI Galaxy Survey is based on observations made at the Anglo-Australian Telescope. The Sydney-AAO Multi-object Integral field spectrograph (SAMI) was developed jointly by the University of Sydney and the Australian Astronomical Observatory. The SAMI input catalogue is based on data taken from the Sloan Digital Sky Survey, the GAMA Survey and the VST ATLAS Survey. The SAMI Galaxy Survey is supported by the Australian Research Council Centre of Excellence for All Sky Astrophysics in 3 Dimensions (ASTRO 3D), through project number CE170100013, the Australian Research Council Centre of Excellence for All-sky Astrophysics (CAASTRO), through project number CE110001020, and other participating institutions. The SAMI Galaxy Survey website is http://sami-survey.org/ 

Funding for the Sloan Digital Sky 
Survey IV has been provided by the 
Alfred P. Sloan Foundation, the U.S. 
Department of Energy Office of 
Science, and the Participating 
Institutions. 

SDSS-IV acknowledges support and 
resources from the Center for High 
Performance Computing  at the 
University of Utah. The SDSS 
website is www.sdss.org.

SDSS-IV is managed by the 
Astrophysical Research Consortium 
for the Participating Institutions 
of the SDSS Collaboration including 
the Brazilian Participation Group, 
the Carnegie Institution for Science, 
Carnegie Mellon University, Center for 
Astrophysics | Harvard \& 
Smithsonian, the Chilean Participation 
Group, the French Participation Group, 
Instituto de Astrof\'isica de 
Canarias, The Johns Hopkins 
University, Kavli Institute for the 
Physics and Mathematics of the 
Universe (IPMU) / University of 
Tokyo, the Korean Participation Group, 
Lawrence Berkeley National Laboratory, 
Leibniz Institut f\"ur Astrophysik 
Potsdam (AIP),  Max-Planck-Institut 
f\"ur Astronomie (MPIA Heidelberg), 
Max-Planck-Institut f\"ur 
Astrophysik (MPA Garching), 
Max-Planck-Institut f\"ur 
Extraterrestrische Physik (MPE), 
National Astronomical Observatories of 
China, New Mexico State University, 
New York University, University of 
Notre Dame, Observat\'ario 
Nacional / MCTI, The Ohio State 
University, Pennsylvania State 
University, Shanghai 
Astronomical Observatory, United 
Kingdom Participation Group, 
Universidad Nacional Aut\'onoma 
de M\'exico, University of Arizona, 
University of Colorado Boulder, 
University of Oxford, University of 
Portsmouth, University of Utah, 
University of Virginia, University 
of Washington, University of 
Wisconsin, Vanderbilt University, 
and Yale University.
\section*{Data Availability}
The data used for this work has been presented in \cite{fraser21} and \cite{woo19}. The SAMI data cubes and value-added products used in this paper are available from Astronomical Optics’ Data Central service at: \url{https://datacentral.org.au/}. The MaNGA data products are available at: \url{https://www.sdss.org/dr15/manga/manga-data/data-access/}.

\section*{Author Contribution Statement}
LC devised the project and drafted the paper. AMF and JW performed the measurements of stellar kinematics and stellar surface density, respectively. LC, AMF, JW, BC and KH contributed to the initial data analysis and preliminary interpretation of the results. JvdS, JBH, JJB and SC provided key support to all the activities of the SAMI Galaxy Survey ('builder status'). All authors discussed the results and commented on the manuscript.






\onecolumn
\appendix
\section{Effect of beam-smearing and inclination corrections on stellar spin}
\label{appA}
\begin{figure}
\centering
	\includegraphics[width=14.1cm]{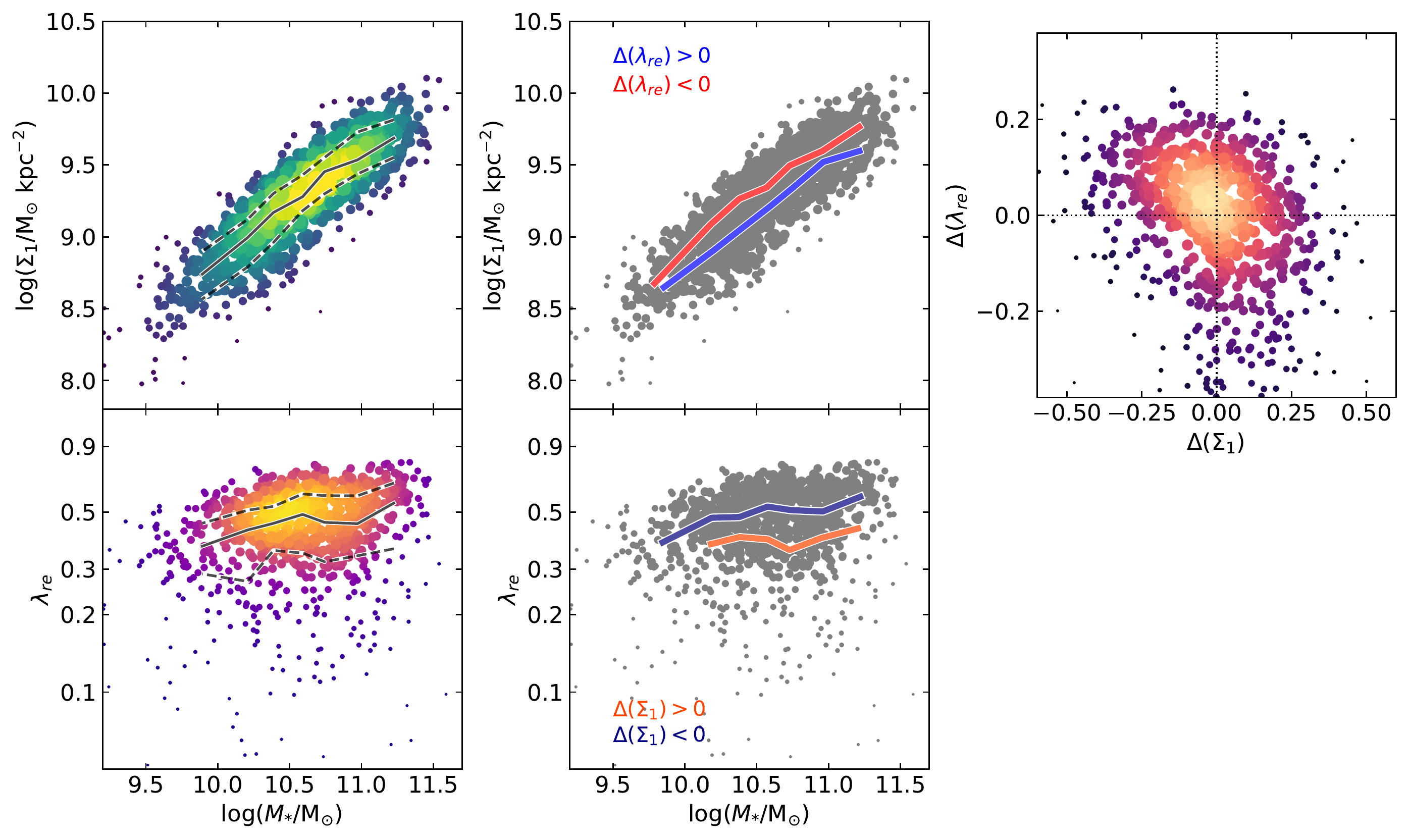}
   \caption{Same as Fig.~\ref{fig:s1_ms}, but with the observed $\lambda_{re}$ plotted instead of the value corrected for both inclination and beam smearing.}
    \label{fig:ms_testcorr}
\end{figure}

\begin{figure}
\centering
	\includegraphics[width=14.1cm]{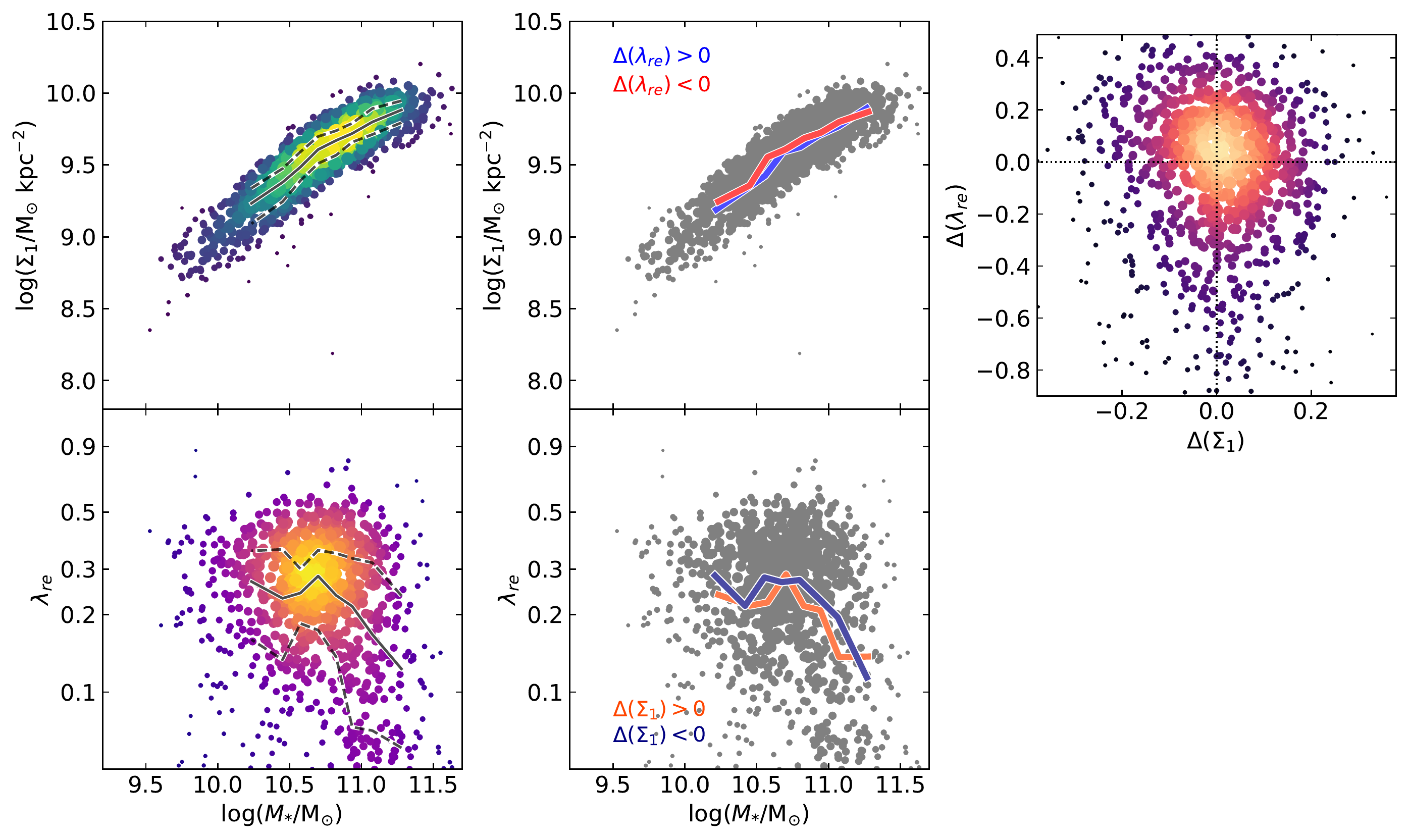}
   \caption{Same as Fig.~\ref{fig:s1_noms}, but with the observed $\lambda_{re}$ plotted instead of the value corrected for both inclination and beam smearing.}
    \label{fig:noms_testcorr}
\end{figure}

In the main body of this paper we have used $\lambda_{re}$ values corrected for both inclination and beam smearing as a proxy for stellar spin parameter ($\lambda_{re}^{c}$). As extensively discussed in the literature (e.g., \citealp{graham18,harborne20,fraser21,vandesande21}), for IFS surveys such as SAMI and MaNGA these corrections combined can easily change the value of $\lambda_{re}$ by $\sim$50\% or more. Thus, it is important to make sure that the main conclusions of this paper are not driven by potential systematic effects in the correction themselves, but are still visible in the original measurements. With this aim, in Fig.~\ref{fig:ms_testcorr} and \ref{fig:noms_testcorr} we reproduce Fig.~\ref{fig:s1_ms} and \ref{fig:s1_noms} by using the observed value of $\lambda_{re}$ instead of $\lambda_{re}^{c}$. As expected, values of spin parameter are lower and the scatter in the $\lambda_{re}$-M$_{*}$ relations for both active and passive galaxies increase (average IQR $\sim$0.23 and $\sim$0.45 dex, respectively). This is due to the large variety of inclinations and observing conditions for the two samples. However, the lack of correlation between $\lambda_{re}$ and $\Sigma_{1}$ ($\rho\sim-$0.09$\pm$0.04) and larger scatter in the $\lambda_{re}$-M$_{*}$ relation for passive galaxies with respect to galaxies on the SFMS still remains. As such, we can conclude that our results are not qualitatively affected by the beam smearing and inclination corrections applied in this work.

\newpage
\section{Effect of seeing on the estimate of $\Sigma_{1}$}
\label{appS1}
\begin{figure}
\centering
	\includegraphics[width=10.5cm]{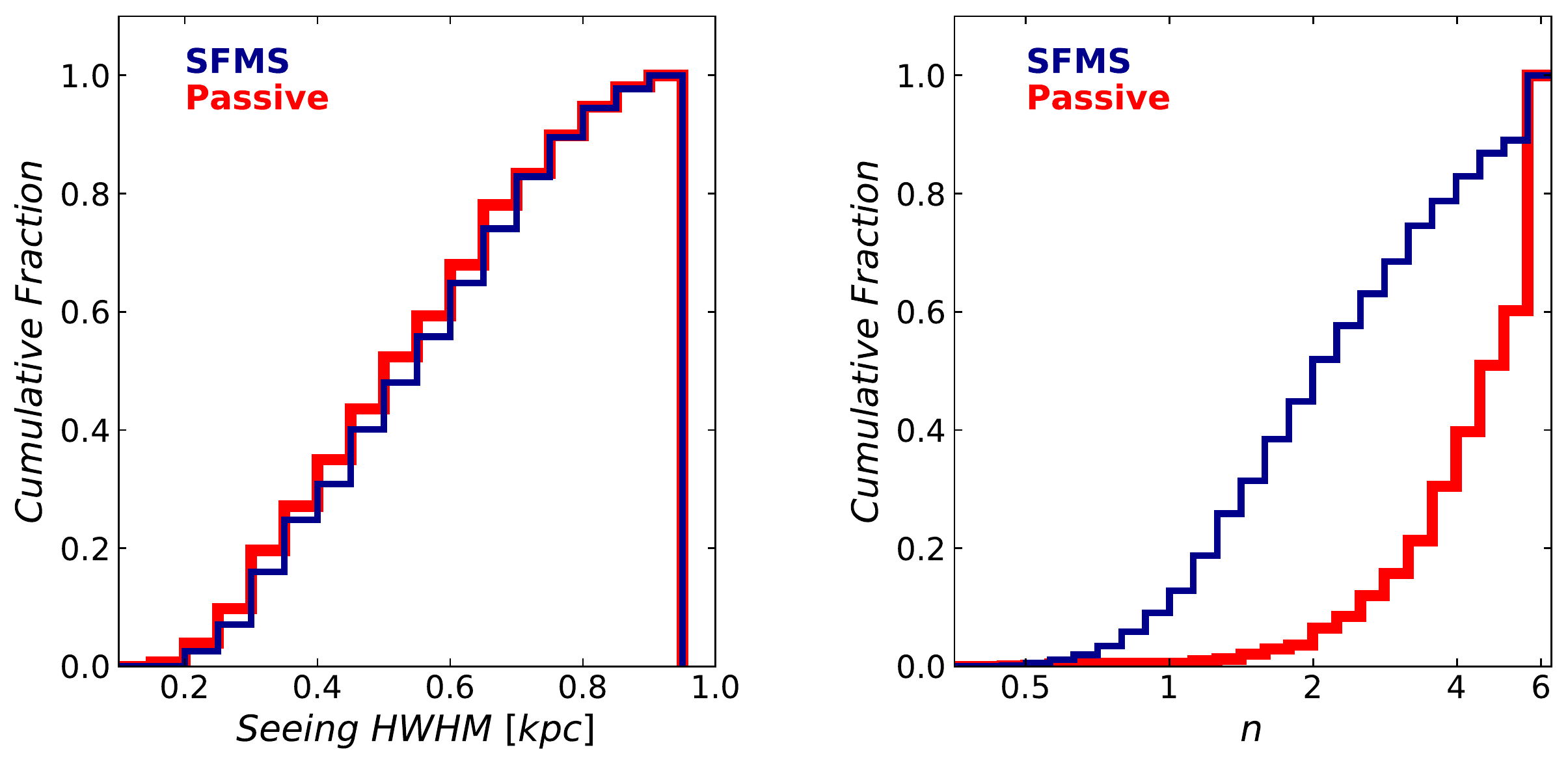}
   \caption{Cumulative distributions of seeing half-width-at-half-maximum (HWHM) in physical units (left) and Sersic indices (right) for SFMS (blue) and passive galaxies (red).}
    \label{fig:s1seeing}
\end{figure}

$\Sigma_{1}$ can be affected by seeing and, while our sample selection ensures that the half-width-at-half-maximum (HWHM) of the SDSS observations is always smaller than 1 kpc, it is important to briefly discuss if this might affect our main conclusions. Indeed, \cite{barro17} have shown that, even when the nominal resolution of the data is 0.6-0.7 kpc,  $\Sigma_{1}$ could be systematically underestimated and that this effect is stronger for galaxies with higher Sersic index. Thus, the lack of correlation between $\lambda_{re}^{c}$ and $\Sigma_{1}$ in passive galaxies could just be the by-product of a larger effect of the atmospheric smearing in these galaxies. Unfortunately, empirical corrections for the effect of seeing applicable to our sample are not available. So, we cannot perform the same test presented in Appendix \ref{appA}. Nevertheless, we can use the work of \cite{barro17}, combined with the properties of our sample, to discuss the potential role of seeing in our findings. In Fig.~\ref{fig:s1seeing} we show the cumulative distributions of HWHM for the SDSS data used to estimate $\Sigma_{1}$ in physical scales (left) and of Sersic index (right) for our SFMS (blue) and passive population (red). Passive galaxies have the same HWHM distribution as SFMS objects and, if anything, they are marginally better resolved. In addition, as expected, passive systems cover a significantly narrower range of Sersic indices than active galaxies. 

According to \citet[see their Fig. 14]{barro17}, at fixed HWHM, the effects of seeing on $\Sigma_{1}$ correlate strongly with Sersic index, so that a population covering a narrow range of Sersic indices will be all affected in a similar way, whereas for a sample spanning a wide range of Sersic indices the changes in $\Sigma_{1}$ may vary significantly (up to $\sim$0.3 dex in the most extreme cases) from galaxy to galaxy. As such, given that our SFMS spans a significantly larger range of Sersic indices than the passive population, we should expect this family to be potentially more affected by atmospheric blurring and any secondary trends to be washed away. Conversely, we still find secondary trends for the SFMS population, making it unlikely that the lack of correlation between $\Sigma_{1}$ and $\lambda_{re}^{c}$ for passive galaxies is just an artefact of the limited spatial resolution of our data (see also the discussion of seeing effects in \citealp{woo15}). Lastly, we note that we have also tested our results by focusing on a sub-sample with HWHM$<$0.5 kpc and our results are unaffected but, due to the significantly smaller number statistics, the quantification of trends as a function of mass and $\Delta(\Sigma_{1})$ or $\Delta(\lambda_{re}^{c})$ becomes highly uncertain, with only two bins of stellar mass having enough galaxies. This is why, contrary to what done in Appendix~\ref{appA}, we do not show the equivalent of Fig.~\ref{fig:s1_ms} and ~\ref{fig:s1_noms} for this subsample.

\newpage
\section{The $\Sigma_{1}$-$M_{*}$ and $\lambda_{re}^{c}$-$M_{*}$ relations for star-forming main sequence and passive galaxies}
\label{appB}
In Table~\ref{tab:scaling} we present the median values (as well as 25\% and 75\% percentiles) for $\Sigma_{1}$ and $\lambda_{re}^{c}$ per bin of stellar mass. Each bin contains 100 galaxies. As discussed in \S~2, we use Eq.~\ref{sfms} to separate star-forming from passive galaxies, labelling as passive all systems below $-$0.5 dex from the locus of the SFMS.

\begin{table}
\caption {The median  $\Sigma_{1}$ and $\lambda_{re}^{c}$ per bin of stellar mass for SFMS and passive galaxies. Uncertainties indicate the 75\% and 25\% percentiles.}
\label{tab:scaling}
\[
\begin{array}{ccc}
\hline\hline
\noalign{\smallskip}
\noalign{\smallskip}
\log(M_{*}/\rm M_{\odot}) &   \log(\Sigma_{1}/\rm M_{\odot}~kpc^{-2})  & \log(\lambda_{re}^{c})  \\                  
\noalign{\smallskip}
\hline
\multicolumn{3}{c}{\rm Star-forming~ main~ sequence~ galaxies}\\
\noalign{\smallskip}
9.89  & 8.74^{+0.16}_{-0.17} & -0.20^{+0.05}_{-0.09} \\[0.1cm] 
10.21 & 8.99^{+0.13}_{-0.20} & -0.17^{+0.06}_{-0.09} \\[0.1cm] 
10.39 & 9.17^{+0.13}_{-0.21} & -0.15^{+0.04}_{-0.08} \\[0.1cm] 
10.59 & 9.28^{+0.15}_{-0.10} & -0.14^{+0.05}_{-0.05} \\[0.1cm] 
10.74 & 9.45^{+0.10}_{-0.15} & -0.14^{+0.05}_{-0.09} \\[0.1cm] 
10.97 & 9.54^{+0.19}_{-0.10} & -0.15^{+0.05}_{-0.06} \\[0.1cm] 
11.22 & 9.69^{+0.13}_{-0.14} & -0.10^{+0.03}_{-0.08} \\[0.1cm] 
\noalign{\smallskip}
\hline
\multicolumn{3}{c}{\rm Passive~ galaxies}\\
\noalign{\smallskip}
10.23 &   9.23^{+0.11}_{-0.14} &  -0.24^{+0.07}_{-0.14}   \\[0.1cm]
10.45 &   9.38^{+0.10}_{-0.13} &  -0.29^{+0.10}_{-0.18}   \\[0.1cm]
10.58 &   9.49^{+0.11}_{-0.09} &  -0.28^{+0.07}_{-0.09}   \\[0.1cm]
10.70 &   9.61^{+0.09}_{-0.10} &  -0.25^{+0.07}_{-0.10}   \\[0.1cm]
10.82 &   9.67^{+0.07}_{-0.10} &  -0.30^{+0.13}_{-0.15}   \\[0.1cm]
10.94 &   9.72^{+0.07}_{-0.06} &  -0.34^{+0.12}_{-0.39}   \\[0.1cm]
11.08 &   9.80^{+0.09}_{-0.09} &  -0.47^{+0.24}_{-0.24}   \\[0.1cm]
11.28 &   9.88^{+0.06}_{-0.09} &  -0.60^{+0.25}_{-0.32}   \\[0.1cm]       
 \noalign{\smallskip}
\hline
\hline
\end{array}
\]
\end{table}


\bsp	
\label{lastpage}
\end{document}